\documentstyle[psfig,rotating,twocolumn]{mn}

\newif\ifAMStwofonts

\newcommand{\bc}{\begin{center}}
\newcommand{\be}{\begin{equation}}
\newcommand{\ee}{\end{equation}}
\newcommand{\ec}{\end{center}}

\def\dd{\, {\mathrm d}}

\def\sp{\epsilon}

\title{Self-similar spherical collapse with non-radial motions}

\author[ Nusser]
      { Adi Nusser
\\ 
Physics Department, The Technion- Israel Institute of Technology, Haifa 32000\\
 E-mail: adi@physics.technion.ac.il}
\begin{document}
\maketitle
\begin{abstract}
We derive the asymptotic mass profile near the collapse center of an
initial spherical density perturbation, $\delta \propto
M^{-\epsilon}$, of collision-less particles with non-radial
motions. We show that angular momenta introduced at the initial time
do not affect the mass profile. Alternatively, we consider a scheme in
which a particle moves on a radial orbit until it reaches its
turnaround radius, $r_*$.  At turnaround the particle acquires an
angular momentum $L={\cal L} \sqrt{GM_* r_*}$ per unit mass, where
$M_* $ is the mass interior to $r_*$. In this scheme, the mass profile
is $M\propto r^{3/(1+3\epsilon)}$ for all $\epsilon >0$, in the region
$r/r_t\ll {\cal L} $, where $r_t$ is the current turnaround radius.
If $ {\cal L} \ll 1$ then the profile in the region $ {\cal L} \ll
r/r_t \ll $ is $M\propto r $ for $\epsilon <2/3$.  The derivation
relies on a general property of non-radial orbits which is that ratio
of the pericenter to apocenter is constant in a force field $ k(t)
r^{n}$ with $k(t)$ varying adiabatically.

\end{abstract}

\section{introduction}

In the hierarchical scenario for structure formation, the initial
density field is Gaussian with a fluctuation amplitude that decreases
with scale (Peebles 1980).  Non-linear gravitational evolution then
causes matter to aggregate into bound virialized objects (halos) which
are believed to harbor galaxies, and galaxy groups and clusters.  The mass
distribution in halos can be inferred from a variety of observations.
These include observations
 of rotation
curves of spiral galaxies (e.g., Persic et. al. 1996), motions of
satellite galaxies (Zaritsky \& White 1994), lensing distortions of
background galaxy images by the potential wells of massive halos
(e.g., Bartelmann \& Schneider 1999), and velocity dispersion and
X-ray maps of galaxy clusters (e.g., Bahcall \& Fan 1998, Sarazin
1986). Nevertheless, no single observational method probes the mass
profile over the entire extent of the halo. Therefore a conclusive
analysis of the observations must rely on an assumed model for the
halo mass profile.  This is important for example in estimating
the total masses of galaxy clusters. Further, halo profiles in the
inner regions depend on the type of dark matter (e.g., Moore 1994).
A comparison between observed and model mass profiles in the context of
a given cosmological model can provide important information on the nature of
dark matter.

We lack a complete theory of nonlinear gravitating systems.  So
a detailed study of nonlinear collapse into bound objects must rely on
N-body simulations.  
The simulations  seem to be converging on
the shape of the mass profile in the halo inner regions
 (Navarro, Frenk \& White 1997,
Moore et. al. 1998,  Jing 1999, Klypin et. al. 2000).  
As of yet no satisfactory analytic explanation of 
 the simulations results has been suggested  (but see Weinberg 2000, 
Syer \& White 1998).
In a generic collapse the infalling matter is in
irregular clumps undergoing dynamical friction and tidal stripping as
they sink towards the center (e.g., Avila-Rees et. al. 1998,  
Nusser \& Sheth 1999). Moreover, a
collapsed object today might have gone through a merger with another
object of comparable mass, a process that might have an effect on its
current mass profile.  Because of these factors, the focus of analytic
studies has been the collapse of perturbations with special
configurations. An important step forward in these studies came with
the realization that the potential in the inner collapse regions
varies very little over the dynamical time scale (Gunn 1977),
admitting an adiabatic invariant for the motion of a particle.  Using
adiabatic invariance, Fillmore \& Goldreich (1984, hereafter FG84, see
also Bertschinger 1985) have shown that in spherical symmetry an
initial density perturbation $\Delta M/M \sim M^{-\sp}$ develops an
asymptotic mass profile $M \propto r$ near the origin for $0<\sp \le
2/3$ and $M\propto r^{3/(1+3\epsilon)}$ for $\sp >2/3$.  FG84
restricted their analysis to particles collapsing on purely radial
orbits. Spherical collapse with non-radial particle motions has also been considered
in the context of adiabatic invariance. Gurevich \& Zybin (1988a,b) developed a
formalism based on the theory of adiabatic capture (e.g., Lifshitz \&
Pitaevskii 1981), to estimate the mass profile in a spherical collapse
including non-radial motions.  Although their formalism can be useful
to study the collapse of various initial conditions, Gurevich \& Zybin
applied it only to the collapse near the center of a smoothed initial
density maxima.  Ryden \& Gunn (1987) included non-radial motions in
their application of adiabatic invariance in a numerical scheme to
study the evolution of a spherical peak in a Gaussian density field
(Bardeen et. al. 1986, Hoffman \& Shaham 1985).

White \& Zartizky (1991) described the $M\propto r$ profile for $\sp
\le 2/3$ in the purely radial spherical collapse in terms of the
crowding of orbits as particles pass through the center.  They
conjectured that if particles had non-radial orbits  then the crowding effect
is avoided and the scaling $M\propto r^{3/(1+3\sp)}$ 
$\epsilon >2/3$ would also be valid for $0<\sp\le 2/3$.  Here we
examine this conjecture in detail by 
generalizing the analysis of FG84.
We consider two schemes for assigning angular momenta to the
collapsing particles. 
 
The outline of this paper is as follows. In section 2 we write the equations of motion and the 
relevant initial conditions and present two schemes for assigning angular momenta
to particles.  In section 3 we discuss self-similarity and adiabatic
invariance in the context of non-radial motions. In section 4 we derive the 
asymptotic profiles for various cases.  We conclude with a summary and 
discussion in section 5.

\section{The equations}
 We consider the evolution of an isolated spherical positive density
perturbation in a flat universe made of collision-less matter only
(density parameter $\Omega=1$).  We describe the perturbation by a
large number of equal mass particles having angular momenta in random
directions such that the mean angular momentum at any point in space
is zero. Spherical symmetry is then preserved and the angular momentum
of each particle is conserved as the system evolves.  It is often
convenient here to think of a particle as a a spherical shell obeying
the equations of motion of a point particle moving in the gravity
field of the perturbation with conserved angular momentum.  We write
now the equations of motion governing the evolution of a shell with a
given angular momentum per unit mass, $L$.  Denote by $r$ be the
distance from the symmetry center, and by $M(<r,t)$ the mass contained
inside $r$ at time $t$.  The trajectory $r(t)$ of the shell as a
function of time is then given by the solution to
\begin{equation}
\frac{\dd^2 r}{\dd t^2}=-\frac{GM(<r,t)}{r^2}+\frac{L^2}{r^3} \; ,
\label{eom}
\end{equation}
where the mass $M(<r,t) $ is determined from the distribution of
shells at time $t$.   Deferring the discussion of the way
shells (particles) are assigned angular momenta, these
equations determine the evolution of the system 
given a set of initial conditions specified at 
an early time $t_i \rightarrow
0$.  We assume that the initial radial velocity of a shell at $r_i$ is
equal to the Hubble expansion velocity at $t_i$, that is,
\begin{equation}
\label{inivel} \frac{\dd r_i}{\dd t}=\frac{2}{3t_i}r_i \; .
\end{equation}
We express  the initial mass distribution in terms of the relative mass excess
$\delta_i\equiv \Delta M/M$ interior to an initial radius
$r_i=[3M/4\pi\rho_b(t_i)]^{1/3}$ where $\rho_b(t_i)=(6\pi G t)^{-2}$
is the mean background density $t_i$.  We take the scale free
form 
\begin{equation}
\label{inid} 
\delta_i=\frac{\Delta M}{M} =\left(\frac{M}{M_0}\right)^{-\sp} \; , \\
\end{equation}
where $\sp>0$. A perturbation with $\sp>1$ can be realized by placing
a point mass at the center of a void with local density contrast
$\propto -M^{-\sp}$ (cf. Chuzhoy \& Nusser 2000 ). 
In cosmology the initial perturbations are small so we restrict the
analysis to shell far enough from the center so that $\delta_i \ll 1$.

The general solution to the equations can be obtained by
 numerical integration
where the mass distribution is continuously updated from the new
positions of the shells at each time step.  However the equations
allow an analytic solution for the motion of a shell that has not
crossed any other so that $M[<r(t)]=const$. We will see later that for
some choices of the angular momenta, this solution is applicable in
the outer regions of the collapse.  The solution can be written
in the  parametric form  (e.g., Landau \& Lifshitz 1976),
\begin{equation}
r=r_0(1-e \cos\eta) ; \qquad t=t_0 (\eta-e\sin \eta) \; ,
\label{nocross}
\end{equation}
where
\begin{equation}
e^2=1+\frac{2EL^2}{G^2 M^2} , \quad r_0=\frac{L^2}{GM}\frac{1}{1-e^2} , \quad
t_0=\sqrt{\frac{r_0^3}{GM}} . \; .
\label{eccen}
\end{equation}
According to this solution, the shell expands until it reaches a
maximum expansion radius (turnaround radius) $r_*=r_0(1+e)$, at time
$t_*=\pi t_0$.
 The solution ceases to
be valid shortly after maximum expansion when the shell crosses other
shells returning from the inner regions after passing through their
maximum expansion sometime ago. The expressions for $r_*$ and $t_*$ 
for purely radial motion can be recovered by taking the limit 
$L\rightarrow 0$ in (\ref{eccen}).  In this limit we find
(cf. Peebles 1980, FG84)
\begin{eqnarray}
r_*&=&r_0(1+e)=\frac{L^2}{GM(1-e)}\\
&\approx& -\frac{GM}{E}=r_i\delta_i^{-1}
=\left[\frac{3M_0}{4\pi \rho_b(t_i)}\right]^{1/3}\left(\frac{M}{M_0}\right)^{1/3+\sp}\\
t_*&=&\frac{\pi}{2 \sqrt{2}} \frac{GM}{|E|^{3/2}}=
\frac{3\pi}{4}\left[\frac{1}{6\pi G\rho_b(t_i)}\right]^{1/2}\delta_i^{-3/2}\\
&=&\frac{3\pi}{4}t_i\delta^{-3/2}=\frac{3\pi}{4}t_i\left( \frac{M}{M_0}\right)^{3\sp/2}  \; .
\label{maxex}
\end{eqnarray}

The solution (\ref{nocross}) also describes the motion of a particle
with non-vanishing angular momentum in an attractive force field
$\propto 1/r^2$.  We remark here that by taking the angular
momentum to zero we do not recover the motion of a particle with zero
angular momentum.  In the limit $L\rightarrow 0$ the particle has zero
radial velocity near the center oscillates between $r=0$ and a maximal
radius, $r_*$.  A particle with $L=0$ on the other hand has an
infinite radial velocity at $r=0$ and oscillates between $-r_*$ and
$r_*$.

\subsection{Two schemes for assigning angular momentum}
We will consider two schemes for assigning angular momenta to
particles. In the first scheme (scheme A) angular momenta are assigned at the
initial time in such a way that no additional scale is introduced in
the collapse.  The angular momentum of each particles is conserved in
the subsequent evolution of the perturbation. In the second scheme 
(scheme B) a
particle acquires its angular momentum when it is at maximum
expansion, prior to that it is assumed to have a purely radial motion.
The angular momentum of a particle in scheme B is $\propto
\sqrt{GM(<r_*) r_*}$ per unit mass (White \& Zartizky 1991), so no
additional physical scale is introduced.

In scheme A the energy per unit mass of a shell can be written
in terms of the initial radius $r_i$ and mass excess $\delta_i M_i$
inside the shell at the initial time, $t_i$,
\begin{equation}
E=\frac{L^2}{2r_i^2}-G\frac{\delta_i M_i}{r_i} \; , \label{eini} 
\end{equation}
so to make all energy components scale similarly we choose
\begin{equation}
L^2=2\alpha G\delta_i M_i r_i=
2\alpha G\left[\frac {3}{4\pi\rho_b(t_i)}\right]^{1/3} 
M_0^{4/3}\delta_i^{4/3-\epsilon}\; ,
\label{scheme1}
\end{equation}
where $1>\alpha >0$ insuring that the energy $E=(\alpha-1)G\delta
M_i/r_i$ is negative and the shell is bound. This choice for $L$ does
not introduce any scale in the initial conditions.

The eccentricity, $e$, corresponding to the motion before shell
crossing is, according to (\ref{eccen}), given by
\begin{equation}
e^2=1+4\alpha(\alpha-1)\delta_i^2 \; .
\end{equation}
Since in cosmological perturbations the initial density contrast is
small, the last relation implies that the eccentricity is very close
to unity. Therefore, according to (\ref{maxex}), the turnaround time
of the shell is $t_*= t_i\delta^{-3/2}\sim r_i^{3\sp}$ and the
turnaround radius is $r_*=r_i\delta_i^{-1}\sim r_i ^{1+3\sp}$. This
scaling means that inner shells collapse earlier than outer shells and
the solution (\ref{nocross}) is always valid outside the the radius of
the shell at maximum expansion at the current time, $t$ (hereafter,
the current turnaround radius).
 
In  scheme B, a shell moves with zero angular momentum until
it reaches its turnaround radius, where it is assigned an angular
momentum per units mass according to
\begin{equation}
L^2={\cal L}^2 G M_* r_*= {\cal L}^2 {G}\left[
\frac{3}{4\pi \rho_b(t_i)}
\right]^{1/3} M_0^{4/3}   \delta^{4/3+\sp} \; , 
\label{scheme2}
\end{equation}
where ${\cal L}$ is a constant value for all shells. To insure that a
shell remains bound and subsequently collapses towards the center we
demand ${\cal L}<1$.  Note that the angular momentum assigned to
a particle at the initial time in scheme A is $\propto
\delta_i^2 M_* r_*$.

Shortly after reaching its turnaround radius, a shell starts its
oscillations in the collective gravitational potential well generated
by shells that have passed their turnaround radii at earlier times.
Angular momentum introduces an effective repellent force, $L^3/r^3$,
preventing shells from reaching the center.  We will see that in 
 scheme A angular momentum affects the density profile only inside
a radius equal to the initial radius of the shell at the current
turnaround radius. For cosmological initial conditions this radius
tends to zero since $t_i\rightarrow 0$. So angular momentum does not
play a role in fixing the density profile of the evolved perturbation.
In this case the results of FG84 for collapse of particles on purely
radial orbits remain valid.  In scheme B, angular momentum can
prevent particles from penetrating a significant fraction, depending
on $\cal L$, of the current turnaround radius.

\section{ Self-similarity and adiabatic invariance}
The initial conditions are scale free, the two schemes for assigning
angular momenta do not introduce any additional scale in the problem.
So the only characteristic scale in the evolution of the perturbation is the
scale of non-linearity which at any time $t$ can be taken as the
 current
turnaround radius, $r_t(t)$.  The mass scale corresponding to the
current turnaround radius is  $M_t=M(<r_t)$.
 So the mass distribution $M(r,t)$
must satisfy the self-similarity condition
\begin{equation}
 M(r,t)=M_t\; {\cal M}(r/r_t) \; , \label{mass:ss}
\end{equation}
where $\cal M $ is a function of $r/r_t$ only. 
Substituting $t_*=t$ in  (\ref{maxex}) we find
\begin{eqnarray}
r_t&=&\left[\frac{3M_0}{4\pi \rho_b(t_i)}\right]^{1/3} 
\left(\frac{t}{t_i} \right)^{2/3+2/(9\sp)}\\
M_t&=&M_0 \left(\frac{t}{t_i}\right)^{2/(3\sp)} \; .
\label{mr:ta}
\end{eqnarray}

Assume that in some region the mass distribution can be approximated
as
\begin{equation}
M=k(t) r^\gamma ,\qquad k(t)=k_0 t^{-s}\; , \label{mpower}
\end{equation}
where $k_0, s$, and $\gamma$ are constants\footnote{For collapse
described by a finite number of particles, the form (\ref{mpower}) is
meant to describe the mass averaged over a few particle oscillations
within $r$.}.  
We will determine $\gamma$ in the next section by generalizing the
analysis of FG84 for collapse with no angular momentum (see also
Zaroubi \& Hoffman 1992).  However, the following useful constraint
\begin{equation}
s+\frac{2}{3\sp}-\frac{2}{3}\gamma\left(1+\frac{2}{3\sp}\right)=0 \; ,
\label{expo:ss}
\end{equation}
can readily be found by substituting the asymptotic form
(\ref{mpower}) for the mass in the self-similarity condition
(\ref{mass:ss}) and using (\ref{mr:ta}).

Throughout the rest of the paper we will assume that the potential
well near the center varies adiabatically (Gunn 1977, FG84). This
means that a shell near the center makes many oscillations before the
potential changes significantly (Gunn 1977, FG84).  In a slowly
varying potential the action variables associated with the motion of a
particle is invariant.  The angular action variable is 
the angular momentum 
which, thanks to spherical symmetry, is conserved  
independent of whether or not the potential changes adiabatically. The radial action
variable associated with the motion of a particle in the potential
well of (\ref{mpower}) is
\begin{eqnarray}
J_r&=&2\pi \int_{r_b}^{r_a}\dd r\left(\frac{\dd r}{\dd t}\right)\\
&=&\int \dd r \left[2\left(E-\frac{\kappa(t)}{\gamma-1}r^{\gamma-1}\right)
-\frac{L^2}{r^2}\right]^{1/2} \; ,
\end{eqnarray}
where $r_a$ and $r_b$ are, respectively, the apocenter and pericenter
of the particle's orbit.  
We prove now that the invariance of $J_r$ implies that the 
ratio $\xi\equiv r_b/r_a$ is constant with
time.  This important property will simplify the determination of
$\gamma$ in the next section.  We prove it in the following
way. 
The energy $E$ and $\kappa$ can be expressed in terms of $r_a$
and $r_b$ as
\begin{eqnarray}
\frac{2\kappa}{\gamma-1}&=&  -L^2 \frac{
r_a^{-2}-
r_b^{-2}
}{ r_a^{\gamma-1}-r_b^{\gamma-1}}\label{EJ1}\; ,\\
2E&=& -L^2 \frac{
r_a^{-2}-
r_b^{-2}
}{ r_a^{\gamma-1}-r_b^{\gamma-1}}r_a^{\gamma-1}+\frac{L^2}{r_a^2} \; .
\label{EJ2}
\end{eqnarray}
Therefore
\begin{equation}
J_r=2\pi L\int_\xi^1\dd u \left[\left(\xi^{-2}-1\right)
\left(\frac{1-u^{\gamma-1}}{1-\xi^{\gamma-1}}\right)
+\left(1-u^{-2}\right)\right]^{1/2}
\end{equation}
The angular momentum $L$ is conserved, so the invariance of $J_r$
means that $\xi$ is constant.  Since
the only special time in the particle history is its turnaround time $t_*$, the
invariance of $\xi$ implies that
\begin{equation}
\frac{r_{a,b}}{r_*}=\left(\frac {t}{t_*}\right)^q \, ,
\label{rastar}
\end{equation}
near the center.  The index $q$ can be expressed in terms of $\gamma$
and $s$ by using (\ref{rastar} ) in (\ref{EJ1}). This yields
\begin{equation}
s=q(\gamma+1) \, .
\label{sq:exp}
\end{equation}

\section{The asymptotic behavior}
Assuming the asymptotic form (\ref{mpower}) we now turn to estimating
the exponents $\gamma$, $s$, and $q$.  Let $P(r,r_i,t)=\int_r^{r_a}
dr(\dd r/\dd t)^{-1}/\int_{r_b}^{r_a} dr(\dd r/\dd t)^{-1}$ the
fraction of time a particle with initial radius $r_i$ spends inside
the radius $r$ at the present time. The function $P$ is well defined
if the mass within $r$ changes very little during one oscillation
period of a particle inside $r$.  The mass $M(<r,t)$ can be written in
terms of $P$ as
\begin{equation}
M(<r,t)=\int_0^{M_t} \dd M_i P(r,r_i,t) \; ,
\label{mfract}
\end{equation}
where  $M_i=4\pi \rho_b(t_i) r_i^3/3$. 
By substituting (\ref{mass:ss}) and (\ref{mpower}) in this 
last relation we obtain
\begin{equation}
\left(\frac{r}{r_t}\right)^\gamma=\int_0^{M_t}\!\frac{\dd M_i}{M_t}
P(r,r_i) \; .\label{Praw}
\end{equation}
Following FG84 we write $P$ in terms of $u\equiv r/r_a$.  
By definition, $P(u)=0$ for $u<\xi$ and $P(u)=1$ for $u>1$.
For $\xi \le u\le 1$, we write $P(u)=I(u)/I(1)$,
where
\begin{equation}
I(u)= \int_{\xi}^u\! \dd v\left[\left(\xi^{-2}-1\right)
\left(\frac{1-v^{\gamma-1}}{1-\xi^{\gamma-1}}\right)
+\left(1-v^{-2}\right)\right]^{-1/2} .
\end{equation}
Using the relations (\ref{maxex}) and (\ref{rastar}) 
we express the mass, $M_i$,  in terms of $u$. So 
\begin{equation}
\left(\frac{r}{r_t} \right)^{\gamma-p}
=\frac{1}{p}\int^\infty_{r/r_t}{\dd u}u^{-(1+p)}
P(u,\xi) \; ,
\label{Pfinal}
\end{equation}
where 
\begin{equation}
p= \frac{6}{2+3(2-3q)\epsilon} \; .
\label{p:exp}
\end{equation}
Particles with $ r_b> r$, i.e., $u=r/r_a< \xi$, have $P(u)=0$ and do
not contribute to the integral on the right hand side.  The relation
(\ref{Pfinal}) allows us to obtain the asymptotic behavior in scheme A and B
without a detailed calculation of $P(u,\xi)$.
 
\subsection{Scheme A}
The first step in the derivation of the asymptotic behavior is to
relate $\xi$ to the initial density contrast.  This can easily be done
by substituting the expression (\ref{scheme1}) for $L$ in (\ref{EJ1}),
and using (\ref{expo:ss}) and (\ref{sq:exp}).  The result is
\begin{equation}
\frac{\xi^2-\xi^{\gamma+1}}{\xi^2-1}=
-\alpha(\alpha-1)(\gamma-1)\delta_i^{2} \; .
\end{equation}
Since $\delta_i\ll 1 $ this relation implies that $\xi \ll 1$ as well.
Retaining only the lowest order in $\xi$, the relation gives $\xi
\propto \delta_i $ for $\gamma>1$, and $\xi\propto
\delta_i^{2/(\gamma+1)}$ for $\gamma<1$.  Since $r_a\le r_*$, we have
$r_b=\xi r_a\le \xi r_*=\xi r_i\delta^{-1}<r_i$, i.e., the pericenter
of a shell 
is smaller than its  initial radius.
Except near $r_b$, where a shell spends only a tiny fraction
of its orbital time\footnote{ Consider the orbit of a particle
in the plane $(r,\psi)$ where $\psi$ is the angular position.  The
time spent near $r_b$ can be estimated as $\Delta T=
\int_{\psi_1}^{\psi_2}\dd \psi r^2/L < 2\pi r_m^2/L$, where $r_m$ is a
few times $r_b$. By (\ref{EJ1}) we find that $r_b^2/L \sim \xi^n T_c $
where $n>0$ and $T_c$ is the period of a circular orbit at the
apocenter $r_a$.  So the time fraction spent near $r_b$ is negligible
for $\xi\ll 1$.}, the motion  is almost radial.  
The initial radius, $r_{ti}$, of the shell currently at turnaround sets 
an upper limit on $ r_b$. For  $r\gg r_{ti}$ all shells move on almost purely
radial orbits and  the analysis of FG84 for collapse on purely radial
motions apply. Therefore, according to FG84, we have
\begin{equation}
\gamma=p=\frac{3}{1+3\sp} \;,\quad s=q=0 \, ,\qquad {\mathrm for}\quad 
\sp \ge \frac{2}{3}\; ,
\label{gpower}
\end{equation}
and,
\begin{equation}
\gamma=1 \;,\quad q=\frac{3\sp-2}{9\sp},\quad s=2q \, ,
\qquad {\mathrm for}\quad 
\sp < \frac{2}{3}\; .
\label{gpower}
\end{equation}
However $r_{ti}$ tends to zero as $t_i\rightarrow 0$. So for 
cosmological initial conditions in which $t_i\rightarrow 0$ the 
the collapse is identical to that of particles on purely radial motions.

\subsection{ Scheme B}

The angular momentum in this scheme is given by (\ref{scheme2}). So the
relation ({\ref{EJ1}) gives
\begin{equation}
\frac{\xi^2-\xi^{\gamma+1}}{\xi^2-1}=
-{\cal L}^2(\gamma-1) \; ,
\end{equation}
implying that  $\xi$ is the same for all particles.  The radius $r_t \xi$
is an upper limit on the pericenters of all shells. So only particles
which have passed their turnaround radii early enough can contribute
to the density at $r\le r_t\xi$.  
 Particles at $r\ll r_t \xi^2 $ have  their
 apocenters  inside  $r_t\xi$ as well, and so the radius 
$r_t\xi^2$  marks the boundary of the inner region where 
we expect the same  power law behavior for the mass distribution.
 In this region only particles with
$r_b=\xi r_a\le r$, i.e., $u\ge \xi$ contribute to the integral in
(\ref{Pfinal}), so
\begin{equation}
\left(\frac{r}{r_t} \right)^{\gamma-p}
=\frac{1}{p}\int^\infty_{\xi}{\dd u}u^{-(1+p)} P(u,\xi) \; .
\end{equation}
The integral is independent of  $r$, therefore we must have
$ \gamma=p $
for all $\epsilon >0$. Using the expression (\ref{p:exp}), and the 
relations
(\ref{expo:ss}) and  (\ref{sq:exp}), yields
\begin{equation}
\gamma=p=\frac{3}{1+3\sp} \;,\quad s=q=0 \; , \quad {\mathrm for} \quad 
{\mathrm all } \quad \sp>0 \; .
\label{gpower}
\end{equation}
In the collapse with $L=0$ (FG84) and in scheme A
 this result is correct only for $\sp \ge 2/3$.

If $\xi\ll 1$ then in the region $r_t \gg r\gg r_t\xi$ particles move
on almost radial orbits and the analysis of FG84 for collapse with no
angular momentum is applicable.  Applying the analysis of FG84
in that region gives $\gamma=1$ for $\sp\le 2/3$ and $\gamma=p$, as in
(\ref{gpower}), for $\sp>2/3$.  Therefore, as $r$ is increased, the
mass profile index varies from $\gamma=p$ to $\gamma=1$ for $\epsilon
\le 2/3$, but remains $\gamma=p $ for $\sp >2/3$.  In figure 1 we show
the mass profiles obtained from spherical N-body simulations of 12000
shells for $\sp=0.3$ with angular momentum introduced according to scheme B.
 The curves correspond, respectively, to
simulations with ${\cal L}=0.9$ (solid), 0.3 (long dashed), 0.1
(short dashed), and 0.001 (dotted). To save CPU, a shell participates
in the simulation only after it reaches its turnaround radius.  A
fourth order Runge-Kutta time integration is used to integrate the
equations of motion.  The  mass profiles in the figure  behave
as  predicted from our analysis.  
A particle with $ {\cal L}=0.9$ moves in an almost
circular, so that it effectively circles the halo at a radius slightly
smaller than its turnaround radius.  Therefore the corresponding mass
profile (solid curve) would have an index close to
$\gamma=3/(1+3\epsilon)=1.57$ almost for all $r<r_t$. The mass
profile in the inner regions for ${\cal L}=0.3$ and $0.1$ (long and
short dashed, respectively) have $\gamma\approx 1.57$, while in the
outer regions it becomes $\approx 1$. For ${\cal L}=0.001$,
particles move on almost radial orbits over the whole
distance range shown in the plot. Hence $\gamma\approx 1$.

\begin{figure}
\centering
\mbox{\psfig{figure=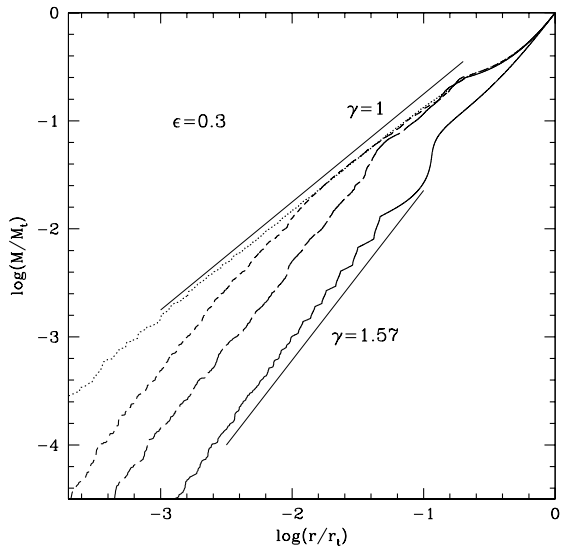,height=3in,width=3in}}
\caption{ Mass profiles from spherical N-body simulations in which
particles acquire angular momenta according to  scheme B.  The
initial density perturbation has $\sp=0.3$.  The solid, long dashed,
short dashed, and dotted curves are, respectively, mass profiles for
${\cal L}=0.9$, 0.3, 0.1, 0.001. }
\label{fig:fig1}
\end{figure}

\section{discussion}
We have studied the effect of non-radial motions on the mass profile
of spherical self-similar collapse.  Assigning angular momentum at the
initial time does not affect the evolved mass profile. This is not
surprising as the initial kinetic energy is dominated by the radial
Hubble expansion velocity $\propto r/t_i$. Any additional finite
velocity is negligible compared to the Hubble expansion velocity as
$t_i \rightarrow 0$. This has also been shown explicitly for collapse
with purely radial motions where a finite radial velocity component is
superimposed on the Hubble expansion at the initial time (e.g.,
Peebles 1980, Padmanabhan 1993).  We have also considered the consequences of a
different scheme (scheme B) for assigning angular momenta to
particles. In scheme B, a particle acquires its angular momentum at
maximum expansion. The acquired angular momentum is chosen in such a
way that the ratio of the pericenter to apocenter is the same for all
particles and no additional physical scale is introduced. In this case
we have confirmed the conjecture by White \& Zaritsky (1991) that the
scaling $M\propto r^{3/(1+3\sp)}$ is maintained for all $\epsilon$ as
$r \rightarrow 0$. But we have also shown that if the added angular
momentum is small, then the behavior $M\propto r$ is restored far
enough from the center for a perturbation with $\sp \le 2/3$.

What is the justification for scheme B in a generic collapse
configuration?  Consider a halo forming in a gravitational collapse of
a high density region in an initial scale free random Gaussian. The
infalling matter is distributed in clumps (satellites).  If the halo
is approximately spherical then the collapse process cannot
introduce any additional scale and angular momentum must be
proportional to the product of the mass and the radius at the epoch of
maximum expansion.  But scheme B also assumes that the angular momentum of
a particle is negligible before it reaches maximum expansion. This is
a reasonable assumption if angular momentum is mainly generated by
tidal interaction between the infalling satellites themselves. Because
the halo is assumed spherical, the halo-satellite tidal interaction is
not expected to generate any angular momentum with respect to the halo
center.  The assumptions underlying scheme B are consistent with
results of N-body simulations (e.g., Barnes \& Efstathiou 1987).
Adiabatic invariance was also applied to non scale free initial
density perturbations (Hoffman \& Shaham 1985, Ryden \& Gunn 1987,
Lokas 2000, Lokas \& Hoffman 2000).  N-body results imply that in
these situations as well, it is reasonable to introduce angular
momentum according to scheme B.

\section{Acknowledgment}
The author thanks L. Chuzhoy and Y. Hoffman for stimulating
conversations.

\end{document}